\documentclass[
]{ceurart}

\usepackage{listings}
\begin{document}
\lstset{ 
  backgroundcolor=\color{white},   
  basicstyle=\footnotesize\ttfamily,        
  breakatwhitespace=true,         
  breaklines=true,                 
  captionpos=b,                    
  deletekeywords={data},            
  extendedchars=true,              
  firstnumber=1,                   
  frame=none,	                   
  keepspaces=true,                 
  keywordstyle=\color{blue},       
  morekeywords={Fact, Var, Bool, Function, Event, Physical, Act, Duty, Identified, Derived, When, Where, Holds, Present, Placeholder, For,
     Actor, Recipient, Holder, Claimant, Related, Violated, Conditioned, Creates, Terminates, Obfuscates,True, False, Extend, externally, Syncs, with, Extends, Open, Closed, Terminated,
     from, when, to, by, Sum, Count, Forall, Exists, Foreach, Predicate,!,+,-,{?},{!?}},            
  numbers=none,                    
  numbersep=1em,                   
  rulecolor=\color{black},         
  showspaces=false,                
  showstringspaces=false,          
  showtabs=false,                  
  stepnumber=1,                    
  tabsize=2,	                   
  title=\lstname                   
}

\copyrightyear{2024}
\copyrightclause{Copyright for this paper by its authors.
  Use permitted under Creative Commons Attribution-ShareAlike 4.0
  International (CC BY-SA 4.0).}

\conference{Jurix'24: AI for Access to Justice Workshop,
  December 11, 2024, Brno, Czechia}

\title{Managing Administrative Law Cases using an Adaptable Model-driven Norm-enforcing Tool}


\author[1]{Marten C. Steketee}[%
orcid=0009-0007-8297-1909,
email=m.c.steketee@uva.nl,
url=https://martensteketee.nl/,
]
\cormark[1]
\fnmark[1]
\address[1]{Informatics Institute, University of Amsterdam, Science Park 900, 1098 XH Amsterdam, The Netherlands}

\author[1]{Nina M. Verheijen}[%
orcid=0009-0005-2728-9064,
email=nina.verheijen47@gmail.com,
url=,
]
\fnmark[1]

\author[1]{L. Thomas van Binsbergen}[%
orcid=0000-0001-8113-2221,
email=ltvanbinsbergen@acm.org,
url=http://ltvanbinsbergen.nl/,
]

\cortext[1]{Corresponding author.}
\fntext[1]{These authors contributed equally.}

\begin{abstract}
Governmental organisations cope with many laws and policies when handling administrative law cases. 
Making sure these norms are enforced in the handling of cases is for the most part done manually. 
However, enforcing policies can get complicated and time consuming with ever-changing (interpretations of) laws and varying cases.
This introduces errors and delays in the decision-making process and therefore limits the access to justice for citizens.
A potential solution is offered by our tool in which norms are enforced using automated normative reasoning.
By ensuring the procedural norms are followed and  transparency can be provided about the reasoning behind a decision to citizens, the tool benefits the access to justice for citizens.

In this paper we report on the implementation of a model-driven case management tool for administrative law cases, based on a set of requirements elicited during earlier research.
Our tool achieves adaptability and norm enforcement by interacting with an interpreter for eFLINT, a domain-specific language for norm specification.
We report on the current state of the case management tool and suggest directions for further development.
\end{abstract}


\maketitle

\section{Introduction}

Digitalisation of governmental services has increased in popularity in recent years, and aims to improve the efficiency and effectiveness of public administration.
%
However, there is not always a direct connection between the digital systems used for these services and the relevant norms.
This disconnection is clearly illustrated by administrative law cases, for which large parts of the decision-making process is done manually, while the communication and registration of cases is automated.
Manually keeping track of the different norms while working on a case is difficult and prone to errors.
A potential solution is provided by automated reasoning about compliance according to formalised representations of norms.
%

%
In this paper we present a case management tool for supporting civil servants with decision-making.
Although the focus in our research was governmental case management, the tool can also be used for other business processes by changing the eFLINT model.
%
%
The norm specification language eFLINT is used to reason about the applicable norms~\cite{VanBinsbergen2020EFLINT:Specifications}.
The current version of the tool is presented in this paper, along with suggestions for further development.
The tool adapts to changes to the (process) model and the norms encoded in the eFLINT specification by changing the user flows of the system.
%
In our work, we apply methods and concepts from business process modelling, model-driven engineering, and normative reasoning.
This paper contributes:
\begin{itemize}
    \item an adaptable case management tool that reasons about applicable norms;
    \item an evaluation against elicited user requirements resulting in suggestions for further development.
\end{itemize}


\section{Background} \label{Background}

\subsection{Normative reasoning/eFLINT} \label{Normative reasoning/eFLINT}
eFLINT is a domain-specific language (DSL) for specifying and reasoning with norms~\cite{VanBinsbergen2020EFLINT:Specifications}.
The language is developed as a DSL to simultaneously specify norms from various sources, as well as (models of) computational processes that can be interconnected with the specified norms.
Given a specification of norms, the eFLINT reasoner can be integrated into systems to check compliance of actions taken in the system.
For the running example, we translated rules for the quittance of municipal taxes\cite{regels-gemeente} into eFLINT code.
This involves adding powers and duties for the different steps in the application process.
To determine whether a client is entitled to quittance of their taxes, the client must provide the public servant with information about their financial position: current income, age, and marital status. 
The model uses eFLINT facts to capture the information of an application and setting thresholds for income. 
In the fragment below, the \lstinline{Var} and \lstinline{Bool} keywords are used to ensure that at most one instance of these types holds true at any given time.
{
\begin{lstlisting}[caption={},label={listing:model}]
Open Var [Age] Identified by Int
Open Bool [Pension age reached]
\end{lstlisting}}\noindent%

The thresholds for capital and income are specified using \lstinline{Var} and are immediately set to a certain value using the following statement.
{
\begin{lstlisting}[caption={}, label={listig:parameters}]
Var [income limit no retirement partner] Identified by Int.
+[income limit no retirement partner](1574).
\end{lstlisting}}\noindent%

The eFLINT reasoner can determine which duties are currently present, and which actions can be taken once the applicant has provided enough information.
This makes sure that the user is informed about compliant actions but the user isn't forced to act in a compliant manner.
The reasoner is able to cope with incompliant behaviour and will show violations of the norms to the user.

The \lstinline{Physical} keyword marks actions which can be executed by users. These synchronise with their institutional counterparts. {
\begin{lstlisting}[caption={},label={listing:acts}]
Act grant-request
  Actor servant
  Recipient applicant
  Conditioned by [Income] <= [Income limit]
  Creates [Quittance granted].

Physical send-approval-letter Syncs with grant-request (servant, applicant)
\end{lstlisting}}\noindent%
%
%

A duty to process the application is created as soon as a client applies for an allowance.
Duties can be associated with violation conditions (one or more Boolean expressions) to determine when a duty is considered violated.
In this example, no violation condition is associated with the duty.

A very important aspect of the eFLINT reasoner is the transparancy the system provides.
The reasoner is able to show the current state of the model and provide a step by step explanation of how this state has been reached.
This enables civil servants and citizens to get a better understanding of the followed procedure and reasoning behind a decision, making it easier to judge wether filing an objection is necessary.

\subsection{User requirements} \label{User requirements}
During earlier research, interviews were held with potential users to gain insight in their requirements for a norm-enforcing case management tool~\cite{automatingCompliance}.
This research was done following a user-centred design method\cite{Martin2012APerspective}.
The seven participants, working at four different organisations, were either involved with handling cases from civilians or organisations subjected to laws and policies, or were involved in the process of turning laws and policies into actions, e.g., within a software application.
The participants that worked for the same organisation had different roles within their organisations. Requirements were elicited from the interview results using a thematic analysis~\cite{Braun2006UsingPsychology}.
These were used as input for the design of the tool.
%
User tests were performed to test the user friendliness of the tool and ensure it was fit for purpose.
Following the user tests, the tool was further developed to demonstrate the ability to reuse the system for different use cases and adapt to changes in norms.
This section will discuss the findings in terms of themes and requirements from the earlier research.


The user requirements were divided in six themes that are laid out and described below:

\begin{itemize}
    \item \textbf{Involved people}: The people that are involved in a case, how many, what role they have, what authorities they have, both on the user side and on the client side.
    \item \textbf{Types of cases}: The types of cases depending on the type of client (e.g., civilian or company), the type of decision made (e.g., request or objection), and possible other factors.
    \item \textbf{Ending of a case}: The ways a case can end, if data from a completed case is stored and if so, for how long.
    \item \textbf{Gathered information}: The types of information gathered during a case, methods, and timing.
    \item \textbf{Actions and duties}: The actions and duties relevant to a case and from what norms or policies they are derived.
    \item \textbf{Interface requirements}: The features and elements expected in a norm-enforcing case management tool.
\end{itemize}
The main findings will be described in terms of themes (in \textit{italic}) and requirements (in \textbf{bold}).

The \textit{involved people} are divided into three categories: clients, internal parties, and external parties. Clients can be civilians, organisations or civil servants.
Internal parties include all involved people from the organisation that take on the case.
Multiple people are often involved to satisfy the `\textbf{four eyes-principle}', stating that two individuals need to approve some action before it can be taken.
%
%
External parties include experts or authorities of different kinds.
%
%
The different people involved do not all have the same level of access, powers, and duties. This means that a form of \textbf{role-based authorization} is required to assign users the correct roles and, as a consequence, the correct normative positions.

The different \textit{types of cases} involve different \textit{workflows}. A general workflow-pattern emerged that can be broken down into: 1) a client asking for a decision on some matter, 2) making a decision based on gathered information, and 3) reporting the decision back to the client. 
The specifics of a use case determine the details of the three steps and is based on multiple factors, such as the type of client and the type of information that needs to be gathered.
For example, whether the client is a civilian or an organisation can influence what information needs to be gathered.

Another factor is the type of decision being requested. 
For example, the decision on an objection has a different process than a (permit) request. 
The amount of case types varies from only one type of case to hundreds of different processes, where each process has slightly different steps.
Therefore, \textbf{workflow models} are required in order for the system to be able to adapt to different types of cases.

The \textit{ending of a case} involves making a decision of some kind. A request can be either approved, denied, or not taken under consideration.
In most cases, an objection to the decision can be made.
If so, steps will be taken to handle the objection, resulting in a decision on the objection.
An objection can thus be seen as asking for a decision related to a previous case.
Therefore, it should be possible to \textbf{recall the details of previous cases}, including closed cases.

The \textit{gathered information} for a case can come from normative sources, such as laws and policies, or can be provided by clients and external parties.
Information comes in all sorts of formats, being online forms, e-mails, phone calls, paper or one-to-one conversations. The most important thing is that the information should be stored.
%
%
One participant mentioned they would rather use the zero-knowledge protocol~\cite{Feige1988Zero-knowledgeIdentity}.
Although information from finished cases is often stored, most participants indicated that they were not certain for how long or under which access conditions.
From this we conclude that recording of information about cases should itself be subject to policies.

The \textit{actions and duties} associated with a case can be divided into those for internal parties and those for the client or an external party.
The client often has a duty to provide requested information.
If this duty is not fulfilled, a decision could be made with incomplete information, which may not favour the client. 
The case manager, on the other hand, has the duty to make a decision, and sometimes within a given period. 
Other actions and duties are specific to the type of case and are written down in relevant laws and policies. 
An example of such an action is informing a client about the decision.

Regarding \textit{interface requirements}, most participants indicated that there are pieces of information they want immediate access to (on a landing page).
Most important are the cases they are currently working on, or depending on their role, certain norms they are working with.
Moreover, the user should be made aware of approaching deadlines.
%
%
Participants mentioned being able to sort and filter is desirable as they are working on a large amount of cases at the same time.
They also indicated they want to see the status of a case as well as their tasks related to the case. 

Regarding the details of a specific case, most participants emphasised the need to have a \textbf{link to relevant sources} of norms and (other) gathered information when working on a case.
Reasons for the requirement include transparency about and the soundness of the decision-making procedure.
The current state of the case should be made clear by showing actions that need to be taken, actions that have been taken, gathered information about the client or case, how this information was provided (e.g., internally, externally or directly by the client), relevant source of norms, and whether any violations of norms occurred.
Most participants indicated that information often comes in batches rather than all at once.
From this we conclude that \textbf{reasoning with partial information} should be supported. In addition, participants reported being interested in \textbf{simulation}, where the idea is to determine the effects of a change in norms or the effects of an action of an ongoing case.


\section{Related work} \label{Related work}
Investments in software systems for governmental services are mostly driven by the need to improve efficiency and effectiveness. Transparency is essential to governmental processes in order for civilians to trust the decisions being made~\cite{vanEngers2019governmental}. There is a need for transparency both by civilians, who want insight in the systems that are used by the government, and by civil servants, who want to ensure systems are correct before they use them for their services. Moreover, there is a need for explainability regarding the decisions made by governmental software, including compliance. As eFLINT applies a form of symbolic artificial intelligence, the application, and effects of rules can be made insightful, for example using argumentation~\cite{calegari2021}.

This paper takes a model-driven approach to attain adaptability and reusability. An overview of DSL technology and model-driven approaches is provided by Rodrigues da Silva~\cite{rodriguesdasilva2015}.
Multiple other languages have been designed to specify norms. Symboleo~\cite{sharifi2020symboleo} and instAL~\cite{Padget2016} are examples of languages that are closely related to eFLINT, as they are also based on the Event Calculus~\cite{kowalski1989logic}. Their event-based nature makes these languages candidates for automating compliance in the way of this paper. Symboleo and eFLINT are both based on the normative concepts of Hohfeld~\cite{hohfeld1917fundamental}. A comprehensive survey on business process compliance is provided by Hashmi et al.~\cite{Hashmi2018}.

In order to keep up with frequently changing tax laws, the Dutch Tax Administration developed the controlled natural language RegelSpraak~\cite{lokin2020regelspraak}, based on the standard Semantics for Business Vocabulary and Rules (SBVR)~\cite{SVBVR} and the RuleSpeak language~\cite{RuleSpeak}. RegelSpraak is simultaneously machine-readable and understandable to both legal experts and IT-developers, by using modeling patterns and language conventions. Since RegelSpraak syntax is similar to the Dutch language, it is easy to understand for users that are not experts in RegelSpraak semantics, increasing explainability. RegelSpraak has been designed to describe computational rules and to handle complex arithmetic and conditional expressions found in tax laws. Unlike eFLINT, Regelspraak is not used to describe powers, rights and duties.

\section{Adaptable, Norm-Enforcing Case Management Tool}
\subsection{Adaptability and reusability} \label{Adaptability and reusability}

To ensure the adaptability and reusability of a norm enforcing application, it is important to make it entirely model-driven. By not storing a structured local copy of the data needed to reason, the tool has to query the reasoner to render the user interface.
Therefore, the developer can both change the user interface and modify the behaviour of the application by changing the model. 
The user interface can be changed by modifying the relevant facts, which are shown to the user. The behaviour of the application can be changed by modifying the condition of an act.
%
The adaptability of the tool is demonstrated by looking at the eFLINT model of the Municipal taxes use case.
The maximum capital and income of an applicant are defined as a \lstinline{Var ... Identified by Int}.
Changing the value of these variables will affect the rights of the applicant and therefore changes the behaviour of the tool.
These values can be changed for existing cases without breaking the tool, as they are stored in the model itself and not in the database.
Another way the behaviour can be changed by altering the model, is by adding to, or changing facts in the model.
Adding physical acts will result in extra buttons being shown to the user.
For example, when adding the physical act of sending a letter to the quittance applicant, an extra button will be presented to the public servant, representing this action.
Changing the \lstinline{Holds when} clause of an act will change wether this act is allowed to executed or not. 
Both kinds of changes can be realised in a modular fashion using the type extension mechanism of eFLINT~\cite{binsbergen2022}.
The tool is reusable in the sense that it can be (re-)used with altogether different model specifications.

\subsection{Design decisions} \label{Implementation and design decision}

The eFLINT language has two options to specify the way in which duties can be terminated by the user. The first option is specifying the acts that terminate the duty in the \lstinline{Terminated by} clause of the duty itself.
This approach follows the CALCULEMUS-FLINT method and can be seen in the code fragment defining \lstinline+process-application+ in section~\ref{Background}~\cite{van2016calculemus}.
Specifying the duty this way makes it easy to determine which actions need to be taken to terminate the end-users active duties, as they are given by the duty itself.
The other way to specify the termination of duties is using the \lstinline-Terminates- clause of the terminating act itself.
To determine which acts terminate a given duty in this case, a form of backward-chaining reasoning is required.
The current implementation of the eFLINT reasoner does not support this.
Therefore, our tool assumes the first method of specifying the termination of duties.
%
The representation of a fact in the interface depends on the domain and fact-type given by the eFLINT reasoner.
When a fact is \lstinline {Identified by} an Integer, the interface will adapt and show a number box instead of a text box.
For Boolean values, the tool presents a set of three radio buttons for the values true, false, and unknown.
%
Together, the actions, duties and facts in the eFLINT specification encode a process model, in which actions by the case manager bring the case to a new state in which other facts and duties hold true and other actions may be enabled or disabled.

Our system is model-driven, in the sense that the eFLINT specification determines which actions are presented to the user (the case manager).
%
%
This is implemented by constantly querying the running eFLINT reasoner whether the displayed acts are enabled. 
For example, when a condition is added, changed or removed, the system automatically reflects these changes by changing the status of the button which represents the physical act accordingly.
A dark blue button is presented when the action is enabled and a red button when it is disabled. 
According to eFLINT semantics, an action that is disabled can still be performed, resulting in a violation being raised.
The user is prompted to confirm they intend to perform a disabled action.
The tool keeps track of which case uses which version of the model, as changing the model for already started cases can lead to unwanted behaviour.
This is achieved by storing different version of an eFLINT model in a separate file and storing a reference to this file in the application database. 
When logging in to the tool, an eFLINT reasoner instance is started for every present case using the corresponding eFLINT model.
Newly started cases always use the model specified by the path in the environment files of the tool.
Because it is uncertain whether the initial eFLINT reasoner stays active, the last state of the eFLINT reasoner is also stored in the database.
When the tool detects a reasoner that is not active anymore, it tries to start a new one using the stored state and model version.
A reference to this new reasoner is stored in the database to keep using the new instance.

\section{Example Case Management Tool} \label{Example case management system}

\begin{figure*}[htb!]
    \centering
    \includegraphics[width=\textwidth]{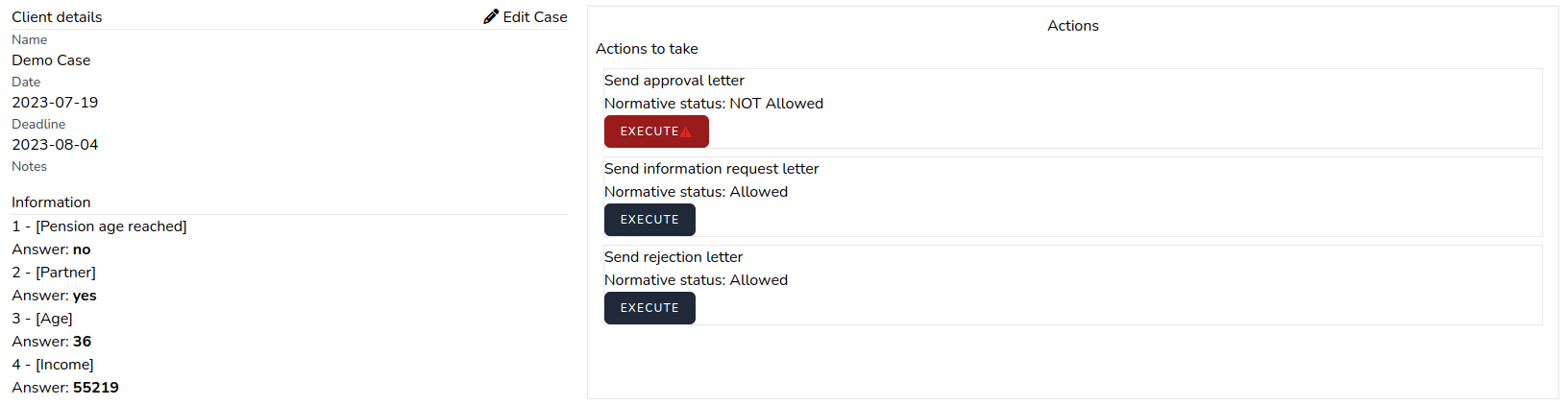}
    \caption{The case overview screen showing the executable actions}
    \label{fig:case_overview_screenshot}
\end{figure*}

\begin{figure*}[htb!]
    \centering
    \includegraphics[width=0.9\textwidth]{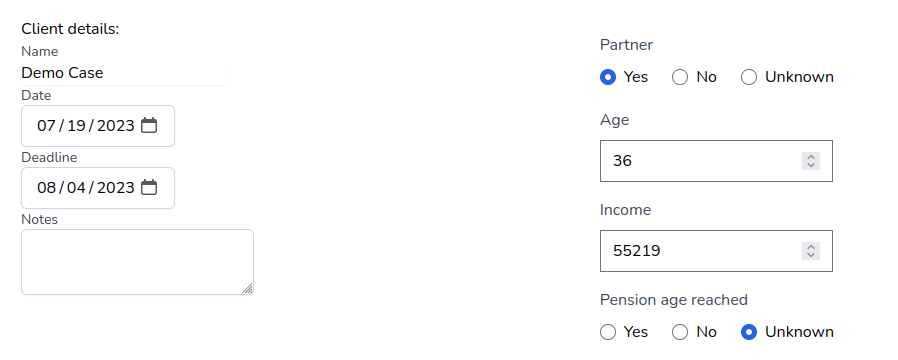}
    \caption{The case editing screen providing inputs for case details.}
    \label{fig:case_editing_screenshot}
\end{figure*}

To demonstrate the features of the tool, we created an eFLINT model describing the process of the quittance of municipal taxes.
This model uses several acts and duties to guide the civil servant through the approval process.
The user interface of the case management tool consists of a case selection screen and two handling screens.
After selecting a case or creating a new case, the public servant is presented with the case details on the left and a set of buttons on the right (figure \ref{fig:case_overview_screenshot}).
The buttons correspond to \textit{physical} action in the eFLINT model, because these physical actions sync with their institutional equivalent, the buttons symbolise certain decisions in the process.
The case details can be changed using a suitable input method.
%

The model-driven approach of our tool ensures that it works with different \textbf{workflows}.
This requires the workflows to be encoded as process models in the eFLINT specification to which the system is applied.
The actions of an eFLINT specification describe a model and normative concepts simultaneously, with pre-conditions (\lstinline-Holds when- and \lstinline-Conditioned by-) determining when an action is compliant and post-conditions (\lstinline-Creates- and \lstinline-Terminates-) determining the power of actors by enabling or disabling actions and duties of (other) actors.
The current design of the system is \textit{action-driven} in the sense that case managers are presented first and foremost with any actions related to cases.
An alternative \textit{duty-driven} design is to present the duties and deadlines related to a case, making the actions that terminate the duties secondary.
In future work, we wish to compare and evaluate implementations of both designs with user studies.

Our system has been designed to handle \textbf{partial information} by allowing information about clients to be collected in a step-by-step fashion.
This is best demonstrated by the `unknown' radio button for Boolean fact-types (figure \ref{fig:case_editing_screenshot}).
The implementation of our tool relies on eFLINT's mechanisms supporting open types for which the closed-world assumption does not apply.
This means that the specification includes types for which the reasoner doesn't need instances to be able to reason with them.

Since the tool is still in development, not all the requirements have been met.
The first user requirement which is not yet implemented is the need to see which legal articles are relevant for the running cases and more specifically, on what legal articles the actions, duties, and violations are based (\textbf{link to relevant sources}). 
We intend to realise this goal by embedding references to the relevant sources in the eFLINT specification, rather than hardcoding them in the system. 
An eFLINT extension is needed, similar to how references to sources are maintained by FLINT~\cite{van2016calculemus}, the language on which eFLINT is based.
Keeping track of source references is slightly more complicated in eFLINT as type definitions can be modularly extended~\cite{binsbergen2022}.
Another implementation that is not yet implemented is the option to give users different roles (\textbf{role-based authorization}).
This enables different levels of authorisation and also enables the option for the \textbf{four eyes-principle}.
Although there is an option to add different roles to the tool, the roles can not yet be used to express policies (such as a policy encoding the four eyes-principle) within eFLINT.
An interaction between the tool and the eFLINT reasoner about user and role information is to be designed in future work.
The simulation system was not implemented in this version of the tool (\textbf{simulation}). However, in other experiments, interfaces for interacting with eFLINT have been designed that at least partially realise the simulation requirement~\cite{frolichDamian2021generic}.
The tool is able to keep track of all information about a case, including after it has concluded (\textbf{recall previous cases}).
However, access to previous cases should be restricted by policies.
More generally, policies are required to regulate the behaviour of the system in ways that are not specific to an individual case.
Access to cases for supervisors of case managers is another example. 
In future work, we intend to demonstrate the enforcement of such policies written in eFLINT within our system.

The forced use of the \lstinline{Terminated by} clause limits the developer, as the developer can only specify the termination of duties looking one step ahead. If the \lstinline{Terminates} clause would have been used, the developer would more easily be able to create a chain of events which terminates the duty, while still being able to present that chain to the end-user. However, since the \lstinline{Terminates} clause requires back-chaining reasoning, which is not yet realised for eFLINT, it was not feasible for the tool.
Another limiting factor is the way the tool handles storing the state of the reasoner.
Because assignments of facts and execution of acts are stored in the database, the developer is limited to adding properties to the model and is unable to change the names of existing facts, acts, and duties for existing cases. 
Removing these attributes could lead to an invalid state being stored in the database and a malfunctioning application. 
In general, a mechanism is required for migrating the knowledge base produced with one version of a specification to a knowledge base suitable for a next version, similar to database migrations.
Although the reasoner is able to show the reasoning behind the decisions made that are made using the tool, the format of this information isn't always comprehensible.

\section{Conclusion}
In this paper, we reported on the implementation of a model-driven case management system for administrative law cases based on a set of elicited requirements.
The underlying eFLINT model assists civil servants in the decision-making process by presenting held duties and the availability of (compliant) actions.
As such, our system automates compliance and has the potential to deliver societal impact by improving the transparency with which governmental services and the laws and policies that govern them are implemented.
This transparency can be found in the way the eFLINT reasoner is able to show the reasoning behind certain actions being compliant and in extension the reasoning behind certain decisions.
The elicited requirements will be used to continue developing our system with additional user tests to (re-)evaluate its design.

\begin{acknowledgments}
  This research is partially funded by the AMdEX-fieldlab project (Kansen Voor West EFRO grant KVW00309), and the AMdEX-DMI project (Dutch Metropolitan Innovations ecosystem for smart and sustainable cities, made possible by the Nationaal Groeifonds)
\end{acknowledgments}

\bibliography{sample-ceur}
\newpage
\appendix
\setcounter{figure}{0}
\makeatletter 
\renewcommand{\thefigure}{A\@arabic\c@figure}
\makeatother
    \section{Overview of the Case Management Tool}
    \begin{figure}[h!]
        \centering
        \includegraphics[width=0.9\textwidth]{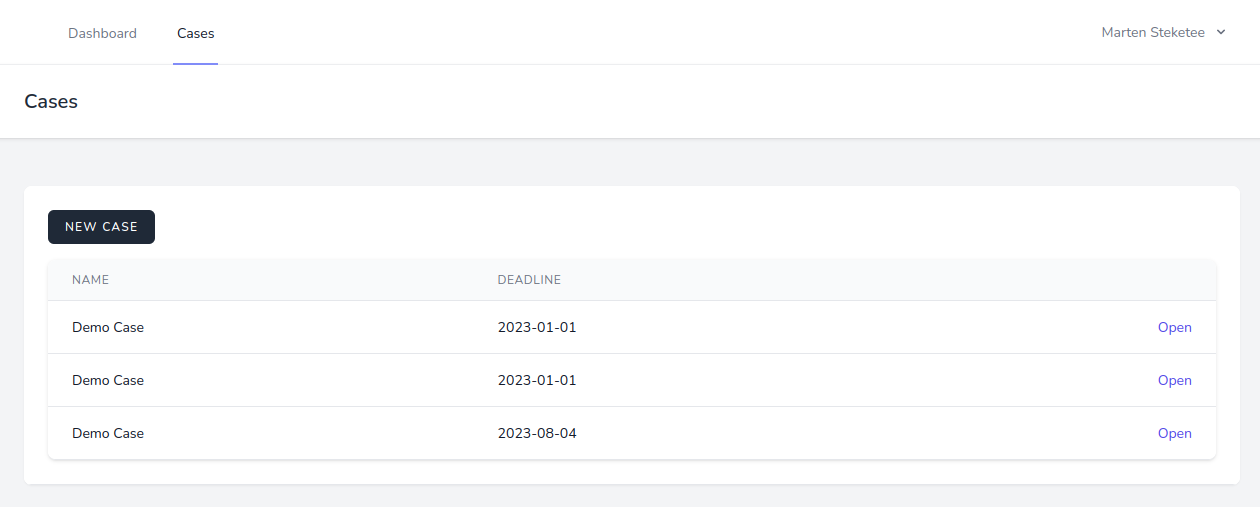}
        \caption{Case selection: After logging in the user can select a case to work on.}
        \label{fig:case-selection}
    \end{figure}

    \begin{figure}[h!]
        \centering
        \includegraphics[width=0.9\textwidth]{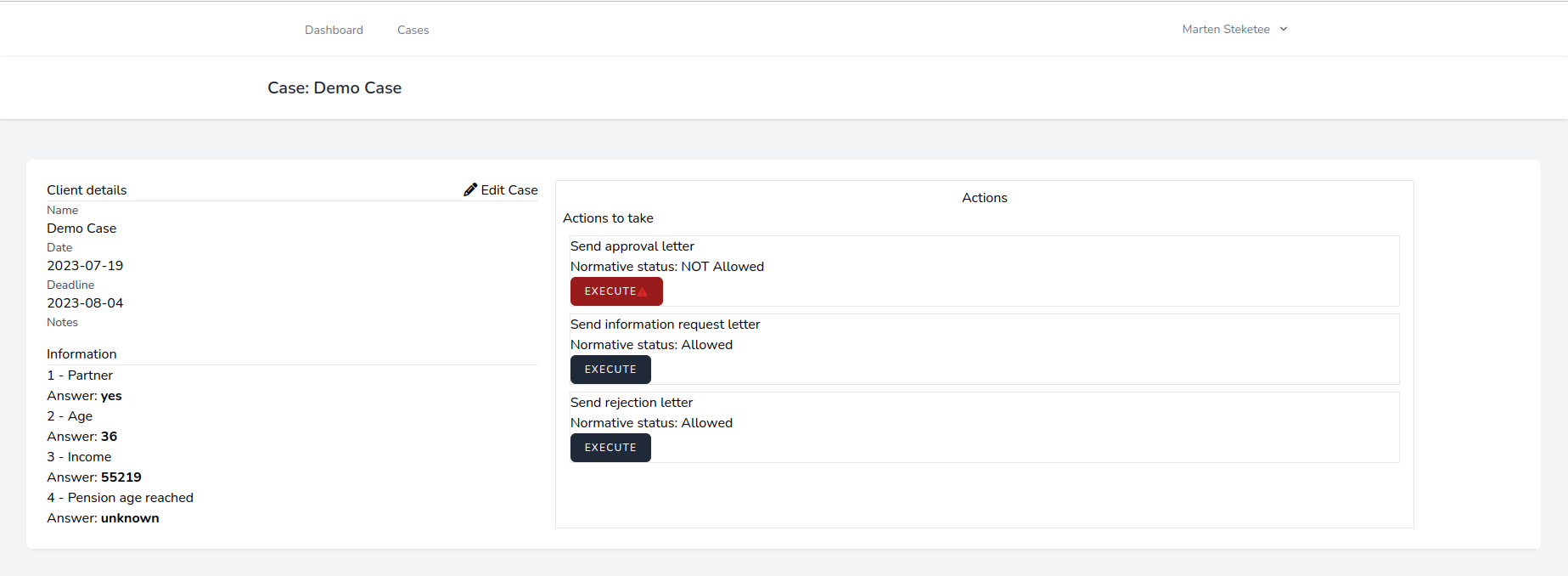}
        \caption{Case overview: When a case is opened the user can view and execute the actions tied to the case. The color of the button represents the normative status. The information which is used to reason about actions is also displayed.}
        \label{fig:case-overview}
    \end{figure}

    \begin{figure}[h!]
        \centering
        \includegraphics[width=0.9\textwidth]{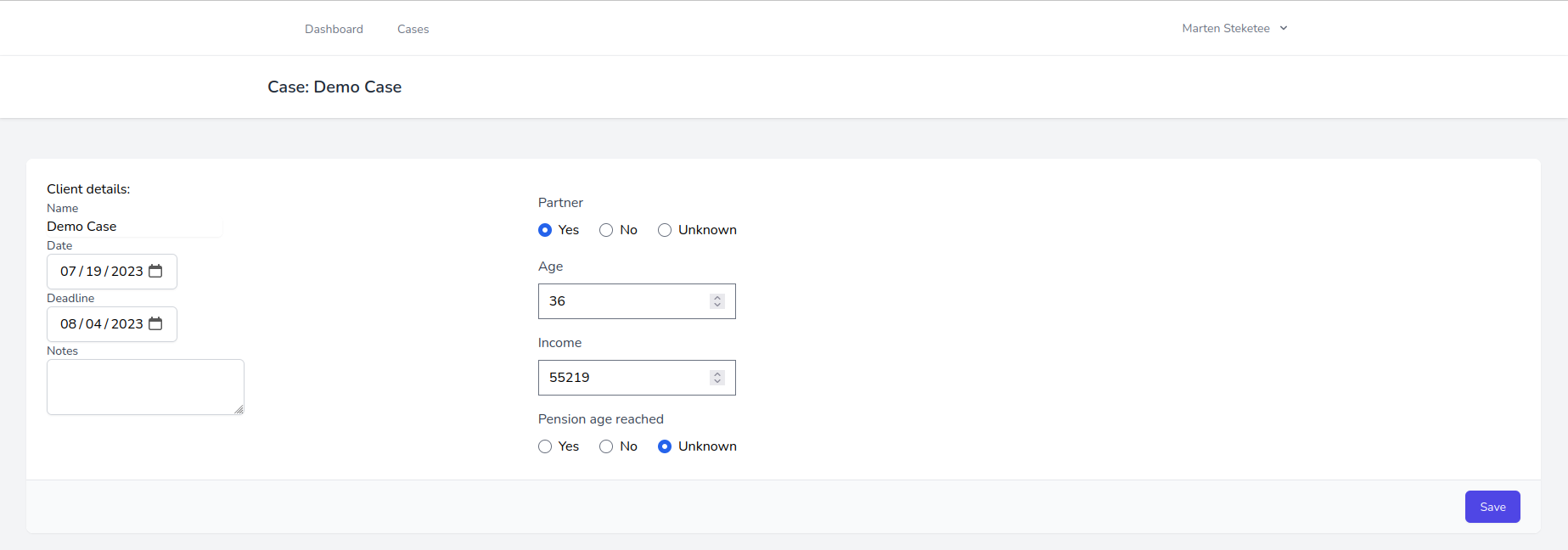}
        \caption{Case editing: To process the information provided by an applicant the user can use the case editing screen. The presentation of the fields is generated using the datatype specified in the eFLINT model. The different types of input are represented by the radio buttons and number fields.}
        \label{fig:case-editing}
    \end{figure}
\end{document}